\documentclass[aps,twocolumn,showpacs,amsfonts,amsmath]{revtex4}

\usepackage{graphicx}  
\let\vec\boldsymbol    
\def\kB{k_{\text{B}}}  

\usepackage{color}

\begin{document}
\title{Glass transition in fullerenes: mode-coupling theory predictions}
\date\today
\author{M.~J.~Greenall}
\author{Th.~Voigtmann}
\affiliation{SUPA, The University of Edinburgh, School of Physics,
JCMB The King's Buildings, Edinburgh EH9 3JZ, U.K.}

\begin{abstract}
We report idealized mode-coupling theory results for the glass transition
of ensembles of model fullerenes interacting via phenomenological
two-body potentials. Transition lines are found for $\text{C}_{60}$,
$\text{C}_{70}$ and $\text{C}_{96}$ in the temperature--density plane.
We argue that the observed glass-transition behavior is indicative of
kinetic arrest that is strongly driven by the inter-particle attraction in
addition to excluded-volume repulsion. In this respect, these systems differ
from most standard glass-forming liquids. They feature arrest that occurs at
lower densities and that is stronger than would be expected for
repulsion-dominated hard-sphere-like or Lennard-Jones-like systems. The
influence of attraction increases with increasing the number of carbon atoms
per molecule. However, unrealistically large fullerenes would be needed to
yield behavior reminiscent of recently investigated model colloids with
strong short-ranged attraction (glass-glass transitions and logarithmic
decay of time-correlation functions).
\end{abstract}

\maketitle

\section{Introduction}

Fullerenes, hollow cages of carbon atoms ($\text{C}_N$), were discovered and
first synthesized in the mid-nineteen-eighties \cite{kroto,kraetschmer}. The
behaviour of $\text{C}_{60}$ has been particularly thoroughly studied.
There are many reasons for the interest in these molecules, notably their
stability and near-spherical structure as single molecules.


Liquids composed of fullerene molecules have long been proposed,
but there are two main fundamental issues concerned with their existence.
Molecular dynamics simulations (Refs.~\cite{wang92,xu,kim,ohno_stability,serra}) predict that individual
fullerene molecules should remain intact up to temperatures of several
thousand $\text{K}$, and arc discharge experiments indicate that they are stable on
at least the $0.1\,\text{ms}$ timescale.  However, when fullerenes are
heated in bulk to relatively low temperatures (around $1200\,\text{K}$), a
dissociation into forms of amorphous carbon sets in (e.g. Refs.~\cite{stetzer,sundar,leifer95,moseler05}). This instability
could be due to impurities \cite{abramo_caccamo_stability}, or be a collective
effect involving inter-fullerene interactions \cite{sundar}.

Secondly, even if the stability of the high-temperature bulk fullerene system
is assumed, there has been considerable disagreement on the characteristics of
its equilibrium phase diagram, in particular the existence or absence of a
thermodynamically stable liquid phase \cite{cheng,hagen}. $\text{C}_{60}$
appears to be a borderline case \cite{ashcroft,mederos}, and the gas--liquid
spinodal eventually shifts into the metastable region below the fluid--crystal
transition with increasing number of carbon atoms per molecule, $N$
\cite{fernandes03,chen03}.
For this reason, model fullerenes have been mentioned as analogues of
short-range attractive systems \cite{bolhuis,rascon}. Such systems can be
realized in colloidal suspensions where added free polymer induces an
effective interaction whose width and strength can be fine-tuned. They are
frequently modeled by a square-well system (SWS), i.e., a hard-sphere (HS)
potential supplemented by a square-well attraction of finite range.  As
this range is decreased, the gas--liquid critical temperature drops, until it
eventually falls into the fluid--crystal coexistence region. Thus the
metastability of the liquid phase can be taken as a signature of
`short-ranged' attractions (see Refs.~\cite{gast,lekkerkerker}). Indeed,
common models for fullerene interaction assume LJ-like interactions between
the individual carbon atoms, which are of short range compared to the large
diameter of the $\text{C}_N$ molecule itself. Furthermore, at high enough
temperatures, the internal degrees of freedom of the fullerene molecules
should not play a major role, since each molecule is able to rotate freely
\cite{yannoni}. In this way, fullerenes suggest themselves as a molecular system that might display the hitherto purely colloidal class of phenomena associated with
short-ranged interparticle attraction.

If one is concerned with the kinetic phenomena of the (perhaps metastable)
fluid, this analogy raises an interesting question. In experiments on
colloid-polymer mixtures \cite{pham,eckert}
and theoretical studies on the SWS or similar models \cite{bergenholtz,dawson},
qualitatively new dynamical features are connected to the short range of the
attractions.
In particular, these systems form glasses; that is, amorphously arrested states,
whose nature changes as the importance of interparticle attraction over
the core repulsion changes. At very high temperatures, the glasses formed
from dense fluids are driven almost exclusively by the core repulsion -- as
particles become trapped in \emph{cages} of their neighbors -- and
are hence qualitatively similar to hard-sphere glasses. Addition of
attractive interactions that are comparable to or bigger in range than the
typical interparticle separation in this high-density fluid has a small
effect on the kinetics, since these add up to a more or less flat background
\cite{widom}. In particular, the glass transition in the Lennard-Jones (LJ)
model (after all a good representation of many molecular liquids) is, according to recent
theoretical studies \cite{tv}, {\em repulsion-driven} over the
entire temperature range. The localization length of the particles
in such glasses is comparable to $10\%$ of the core diameter, in agreement with
the Lindemann criterion for freezing, and little influenced by the range of
the interparticle attraction.
However, if sufficiently short-ranged and strong attractions are present,
a qualitatively different glass is found when the thermal energy becomes
comparable to or lower than the attraction depth, as \emph{bonding} provides
a second particle-trapping mechanism. Consequently, the range over which
particles are localized in such an \emph{attraction-driven} glass scales with
the range of this attraction and can be an order of magnitude smaller than
the one predicted for repulsive glass. Furthermore, their degree of arrest
is much larger; in particular they appear mechanically much stiffer than
comparable standard glasses.
The intriguing question then posed by this analogy is: do fullerenes, under
certain conditions, form attractive glasses, or at least ones that display
some characteristics of the colloidal attractive glasses?

In this paper, we explore theoretically the above analogy on the kinetic level
between the liquid of model fullerenes and short-range-attractive model
colloids.
Recent molecular-dynamics (MD) simulations on model fullerene
liquids \cite{abramo_glass,abramo_glass2,ruberto} have shown that there occurs
a positional glass transition in model fullerene liquids at high density,
connected with a freezing-in of the molecules' translational degrees of freedom,
and not to be confused with the rotational glass
transition occurring in fullerene crystals below room temperature
\cite{gugenberger}.
From the MD data, it was concluded that the positional
glass transition is essentially a repulsion-dominated glass transition
reminiscent of hard spheres. However, this judgement
was based solely on the shape of the transition line $\varrho_g(T)$.
Our study shall provide a theoretical basis for the investigation of such
fullerene glasses, by applying the mode-coupling theory of the glass
transition (MCT) \cite{goetze_rev} to the model potentials for
fullerenes employed in these and similar MD simulations.
MCT has been very successful in describing the details of the
repulsive-to-attractive crossover in the SWS \cite{dawson,bergenholtz}, and
hence should be well suited to investigate the analogous question here. It will allow us to predict not only the qualitative shape of the glass
transition line, but also the strength of the arrest and the form of the
time decay of density fluctuations near to the transition.
The investigation of kinetic effects might also be
interesting for a further understanding of the destabilization of fullerenes
(as far as kinetic cooperative effects are involved).

We start by giving a brief introduction to the fullerene models
used in this study (Sec.~\ref{fullerenes}), and a sketch of MCT and
the calculation of glass-transition lines in the density--temperature diagram
of various fullerenes (Sec.~\ref{mct}). Section~\ref{results} presents our
main results, after which a discussion follows in Sec.~\ref{conclusion}.

\section{Fullerene Models}\label{fullerenes}

For temperatures high enough so that internal modes of the $\text{C}_N$
molecules become unimportant, one can model the fullerene bulk
system by particles interacting with a spherically symmetric pair potential.
A popular model of this kind is due to Girifalco \cite{girifalco}. This
potential can be applied to fullerenes formed from any number $N$ of carbon
atoms; however, the assumption of spherical shape and the neglect of
intramolecular vibrations become more and more questionable the larger the
molecules. The model assumes Lennard-Jones interactions with experimentally
determined parameters between the individual carbon atoms, which are then
integrated over hollow spheres representing the buckyball molecules. This
yields
\begin{eqnarray}\label{girifalco_pot}
  V(r)&=&
    -\alpha\left[\frac{1}{s(s-1)^3}+\frac{1}{s(s+1)^3}-\frac{2}{s^4}\right]\nonumber\\
  & &  +\beta \left[\frac{1}{s(s-1)^9}+\frac{1}{s(s+1)^9}-\frac{2}{s^{10}}\right]
  \,,
\end{eqnarray}
where $r$ is the centre--centre separation of two fullerene molecules of diameter
$d$, and $s=r/d$. The constants $\alpha$ and $\beta$, tabulated in
Ref.~\cite{girifalco}, are proportional to $N^2$. A suitable radius $d(N)$ can
be inferred from experiment and simulation
\cite{girifalco,kniaz,abramo_caccamo,wang_giant}, or from the assumption
$d\propto\sqrt N$ \cite{zubov,abramo_epl}. We will use the Girifalco
potential in our work for $N=60$, $70$, and $96$, the most common low-$N$
fullerenes.

Data from \textit{ab initio} calculations on $\text{C}_{60}$ can be better
fitted by a potential due to Pacheco and Ramalho (PR) \cite{pacheco}.
One combines a long-range van~der~Waals expression with a short-ranged
Morse-potential term,
\begin{subequations}\label{PR}
\begin{align}
  V_{\text{PR}}(r)&=F(r)\times M(r)+\left[1-F(r)\right]\times W(r)\,,\\
  W(r)&=-C_6/r^6-C_8/r^8-C_{10}/r^{10}-C_{12}/r^{12}\,,\\
  M(r)&=M_0\exp[\tau(1-r/d_0)]\left((\tau(1-r/d_0)-2\right)\,,\\
\intertext{where $F(r)$ describes the cross-over between the two,}
  F(r)&=\left(1+\exp[(r-\mu)/\delta]\right)^{-1}\,.
\end{align}
\end{subequations}
$C_6$ and $C_8$ may be determined from density functional theory, whilst
$C_{10}$, $C_{12}$, $M_0$, $d_0$, $\tau$, $\mu$, and $\delta$ are fitting
parameters tabulated in Ref.~\cite{pacheco}.
The PR potential is somewhat softer than the Girifalco potential and appears
superior in a number of predictions, but its application to larger fullerene
molecules is not straightforward. For our discussion of qualitative
features, the PR potential serves to demonstrate
to what extent our results depend on the precise form of the potential
chosen.

\begin{figure}
\includegraphics[width=0.8\linewidth]{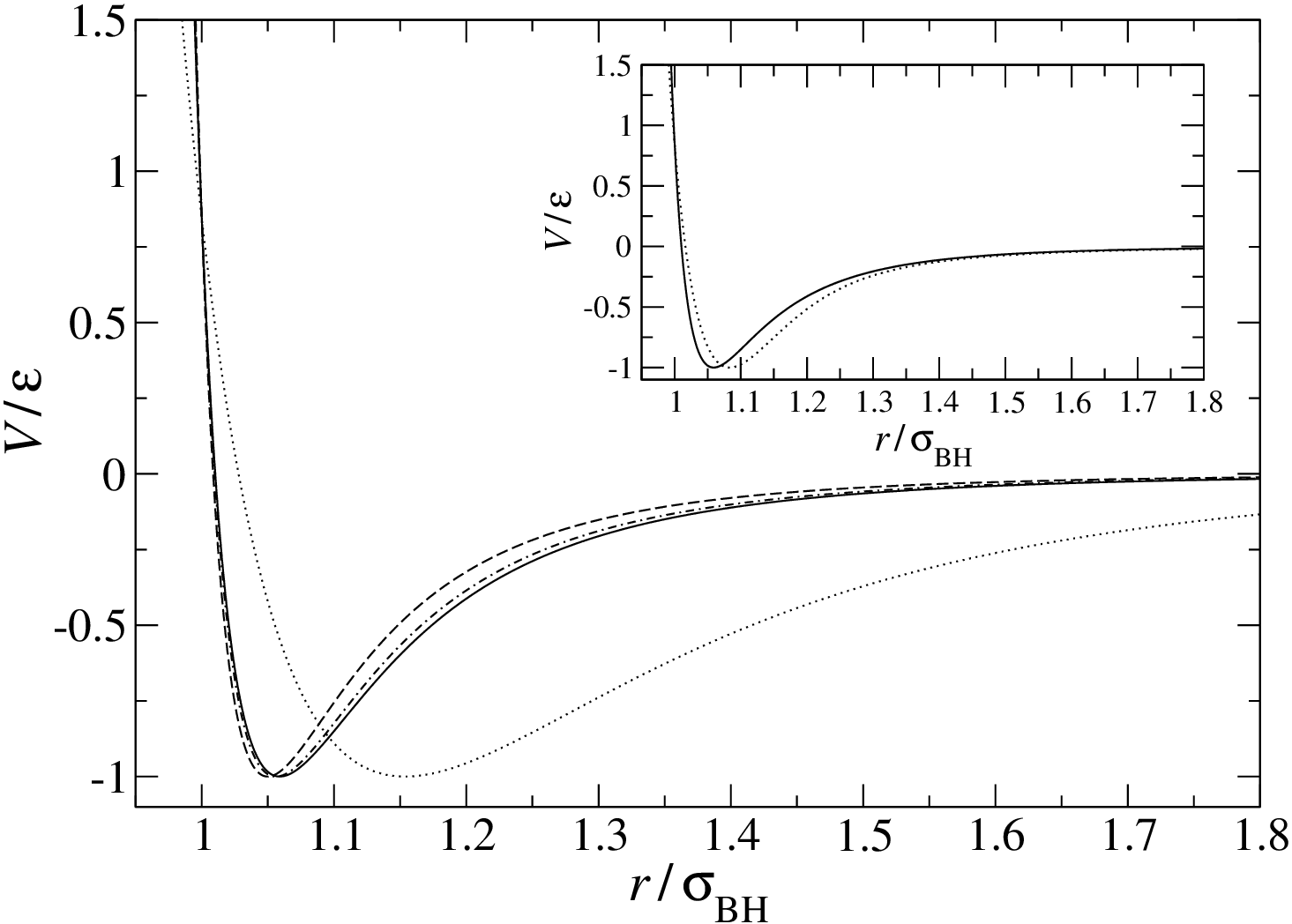}
\caption{\label{potsfig}
  Girifalco potentials for $\text{C}_{60}$, $\text{C}_{70}$, and
  $\text{C}_{96}$ (solid, dash-dotted, and dashed lines), in units of
  reduced temperature and the Barker-Henderson effective diameter
  $\sigma_{\text{eff}}$. The Lennard-Jones potential is shown as a dotted
  line. Inset: comparison of the Girifalco (solid line) and the
  Pacheco-Ramalho (dotted) potentials for $\text{C}_{60}$.
}
\end{figure}

In comparing results for different-sized fullerenes, we need to introduce
suitable natural scales of energy and length. An obvious choice for the
energy scale is $T^*=\kB T/\epsilon$, where $\epsilon$ is the magnitude of
the potential at its minimum. A convenient choice
for the unit of length is given by the Barker-Henderson effective diameter
\cite{barker},
letting $r^*=r/\sigma_{\text{eff}}$ with
\begin{equation}
  \sigma_{\text{eff}}=\int dr\left[1-\exp(-V_{\text{rep}}(r)/(\kB T))\right]
  \,.
\label{sigmaBH}
\end{equation}
$V_{\text{rep}}$ is the repulsive part of the potential only, $V_{\text{rep}}
=V(r)$ for $r\le r_0$, and zero otherwise, where $r_0$ is the point for which
$V(r_0)=0$. As $T\to\infty$, we recover from
this $\sigma_{\text{eff}}\to\sigma$, i.e.\ the hard-core diameter of the
Girifalco potential, and, for hard-sphere systems, $\sigma_\text{eff}\equiv\sigma$. Note however, that the approach to the hard-core
limit is extremely slow: for $\text{C}_{60}$, $\sigma_\text{eff}\approx1.821\sigma$ at the highest temperatures considered ($T=50000\,\text{K}$).
For later comparison we also note that a purely repulsive potential that
incorporates all soft-core effects of the Girifalco potential may be
constructed following the Weeks-Chandler-Andersen (WCA) prescription \cite{wca},
\begin{equation}\label{cutoff-potential}
  V_{\text{cut}}(r)=\begin{cases} V(r)+\epsilon& \text{$r\le r_-$,}\\
    0&\text{$r>r_-$.}\end{cases}
\end{equation}
Here, $r_-$ is the position of the potential minimum. Abramo \textit{et~al.}\
\cite{abramo_glass,abramo_glass2,ruberto}
use this potential to define $\sigma_{\text{eff}}$ according to
Eq.~\eqref{sigmaBH}. We refer to this as
the WCA-$\sigma_{\text{eff}}$ in the following. If one is
interested in discussing changes in the effective range of the potential
with changing $N$, this definition appears disadvantageous, since it includes
part of the change of $r_-/r_0$ in the definition of $\sigma_{\text{eff}}$. 
The approach of this effective diameter to the hard-sphere limit $\sigma$ is even slower than that of the Barker-Henderson definition: for $\text{C}_{60}$, $\sigma_\text{BH}\approx1.827\sigma$ at $T=50000\,\text{K}$. The advantage of $V_\text{cut}$ lies in the fact that it includes the entire section of the potential where $\partial_r V(r)>0$ , and a comparison of MCT results based on $V_{\text{cut}}(r)$ and the full $V(r)$ will be used in the following to separate out those dynamical effects arising from attractive forces.

The Girifalco potentials for $\text{C}_{60}$, $\text{C}_{70}$, and
$\text{C}_{96}$, are shown in Fig.~\ref{potsfig} in units of reduced
temperature and the Barker-Henderson $\sigma_{\text{eff}}$. They show a
weak dependence of the minimum position on $N$,
$r_-^*\sim1/(\text{const.}+\sqrt N)$: $\text{C}_{96}$ might therefore be
expected to display more evidence of attraction-driven glassy behavior
than $\text{C}_{60}$. Even for $N=60$, the minimum is
significantly more short-ranged than in the LJ potential representative for
standard liquids, shown as a dotted line for comparison.
Also shown (in the inset) is the PR $\text{C}_{60}$ potential. The softness of the repulsive component of the PR interaction means that its Barker-Henderson diameter is smaller than that of the Girifalco $\text{C}_{60}$ system, and the potential is shifted to higher $r/\sigma_\text{BF}$.

To apply MCT to any of the above potentials, one requires the
equilibrium static structure factor, $S(q)$, corresponding to $V(r)$.
The liquid-state theory to approximate $S(q)$ is unconnected to MCT. For the
Girifalco potential, we solve the Ornstein-Zernike (OZ) equation numerically to determine
the direct correlation function $c(q)$ (and with it $S(q)=1/(1-\rho c(q))$), using the hypernetted-chain (HNC) and Percus-Yevick (PY) closure
approximations \cite{hansen_mcdonald}.
These both have a known problem termed thermodynamic
inconsistency: two equivalent routes to calculate the equation of state from
them give different results. Much effort has been devoted to improving upon
this deficiency. However, for an application of MCT, this issue is usually
unimportant. As we shall explain below, the MCT integrals are not sensitive to
the $q\to0$ region of $c(q)$, perhaps with the exception of a narrow region
directly surrounding the spinodal of the model. Instead, they rely on a
reasonable description of the $q\gtrsim q_{\text{peak}}$ region, where
$q_{\text{peak}}$ is the first sharp diffraction peak. Similarly, the precise
position of the equilibrium phase-transition lines, and even whether the
approximation predicts a stable or only meta-stable liquid, is not important
here, other than by typically setting an overall shift in the temperature
scale. We have chosen HNC and
PY to indicate boundaries between which many more refined closure schemes will
vary our results \cite{caccamo_rev}. In particular, most of our discussion is based around the PY
approximation, since we found it to give more reasonable results when
probing the hard-core part of the potential at very high temperatures.
The numerical algorithm we employ to calculate $S(q)$ is due to Lab\'{i}k {\em et al}
\cite{sqcalc}; we used a wave-vector grid of $4096$ points with a cutoff
$Q=367.6\,\text{nm}^{-1}$ ($R=35\,\text{nm}$ for the radial distribution function $g(r)$). Our calculated $S(q)$ (using both PY and HNC closures) for $T=1900\,\text{K}$, $\rho=0.74\,\text{nm}^{-3}$ compares well (outside the $q\to 0$ region discussed above) with molecular dynamics data for the corresponding state point published by Alemany {\em et al} in Ref.~\cite{alemany}.

\section{Mode-Coupling Theory of the Glass Transition}\label{mct}

Let us present a brief outline of the (idealized) MCT description of
structural glass transitions. For in-depth reviews, the reader is
referred to Refs.~\cite{goetze_rev,goetze_exp,cummins}.

MCT aims to describe the dynamical arrest that happens in dense liquids:
even in the absence of a thermodynamic phase transition, the viscosity and,
more generally, the relaxation times for many time-dependent fluctuations
in the system increase sharply. As the glass transition is reached, the
fluctuations are no longer able to decay to zero (as it is the case in a
liquid), but a certain amount remains ``frozen'' in the still amorphous
system. The theory describes this as a feedback mechanism driven by slow
density fluctuations.

The most convincing evidence for the MCT scenario has come from colloidal
systems, although the theory deals with both colloidal and molecular liquids
alike. In the latter, however, the MCT transition is ``avoided'' due to
additional relaxation paths called hopping processes and not captured
in the theory. The MCT transition line $T_c(\rho)$, or equivalently
$\rho_c(T)$, can still be inferred from
experimental data by means of scaling laws the theory provides for the
dynamics close to its idealized transition. However, due to the presence of
hopping, the calometric glass transition $T_g$ will be lower, so that $T_g<T_c$ or
equivalently $\rho_g>\rho_c$. The MCT transition point still
indicates a change in transport mechanism from liquid-like to solid-like,
although not a complete arrest.
In this paper, we are concerned not with a precise determination of the
MCT transition point, but rather with its qualitative changes upon varying
control parameters such as temperature $T$ or density $\rho$, and with
the spatial structure of the resulting amorphous solid. These features of
the glass are usually captured quite well by MCT \cite{goetze_exp}.

The structure of the glass shall be characterized by its Debye-Waller factor
(also called form factor or nonergodicity parameter, NEP),
$f^c(q)$: it quantifies the degree of arrest
of density fluctuations at a wave vector $\vec q$ in the glass. It is given
by $f^c(q)=\lim_{t\to\infty}\phi^c(q,t)$, the long-time limit of the
normalized density autocorrelation function,
$\phi(q,t)=\langle\varrho(\vec q,t)^*\varrho(\vec q,0)\rangle/\langle
|\varrho(\vec q,0)|^2\rangle$, at the transition. Here, $\varrho(\vec q,t)
=\sum_{k=1}^N\exp[i\vec q\cdot\vec r_k(t)]$ are the (Fourier-transformed)
density fluctuations, $\vec r_k(t)$ are the individual particle positions,
and $\langle\cdot\rangle$ denotes a canonical average.
$\langle|\varrho(\vec q,0)|^2\rangle=S(q)$ is the static structure factor
of the system.
Even on the liquid side of the glass transition and in presence of hopping
processes, $f^c(q)$ can be determined from the time-dependent density
correlation function. $\phi(q,t)$ (as a function of $\log t$) close to $T_c$
develops
a two-step process consisting of a ``fast'' decay towards its ``plateau
value'' $f^c(q)$, followed by a slow final decay to zero. The time window
over which $\phi(q,t)$ stays close to its plateau extends as $T_c$
is reached, and similar plateaus occur in many dynamical
quantities. The mean-squared displacement (MSD) of a tracer particle for
example exhibits a plateau that is directly connected to the localization
length $r_s$ of this tracer. Also the width of the $f^c(q)$-versus-$q$ curve
gives an indication of the inverse localization length in the glass, since it
closely follows the plateau value for the tagged-particle density correlator,
$f^{s,c}(q)\approx\exp[-q^2r_s^2]$, where the latter approximation holds for
small $q$.
Attractive glasses are characterized by a significantly reduced localization
length, and consequently
exhibit $f^c(q)$-versus-$q$ curves that follow a much broader
envelope than that found in standard hard-sphere or LJ glasses.

Within MCT, the equation determining $f^c(q)$ reads
\begin{equation}\label{mctf}
  \frac{f(q)}{1-f(q)}=m[f](q)\,,
\end{equation}
to be evaluated at the transition point. $m$ is the MCT memory kernel,
\begin{subequations}
\begin{equation}\label{mctm}
  m[f]=\frac{\rho}{2q^2}\int\frac{d^3k}{(2\pi)^3}
    V(\vec q,\vec k)f(k)f(p)\,,
\end{equation}
with $\vec p=\vec q-\vec k$. The $V(\vec q,\vec k)$ constitute the coupling
constants of the theory, and they are determined completely in terms of
the equilibrium static structure of the system; more precisely in terms
of the direct and the triplet correlation function. Lacking a reasonable
expression for the latter, one usually drops it (approximating three-point
averages by a convolution approximation; this is not connected to the
neglect of three-body terms in the potential). One then gets
\begin{equation}\label{mctv}
  V(\vec q,\vec k)=S(q)S(k)S(p)\left[\frac{\vec q\cdot\vec k}q c(k)
  + \frac{\vec q\cdot\vec p}q c(p)\right]^2\,.
\end{equation}
\end{subequations}
In this sense, $S(q)$ is the sole input to MCT. In particular, temperature
effects enter only through their effect on $S(q)$. The $k$-integration in
Eq.~\eqref{mctm} is responsible for suppressing finite variations in the form
of $c(q\to0)$, i.e., the influence of thermodynamic effects.

Eq.~\eqref{mctf} in general has many solutions $\tilde f(q)$; the NEP
$f(q)$ is the largest positive, real of these. As the control parameters
and thus the coupling constants $V$ are varied smoothly, bifurcations in
Eq.~\eqref{mctf} can occur, leading to a (generally) non-smooth jump for
$f(q)=0$ to $f(q)>0$. These bifurcations are the MCT glass transitions.
They are found by a bisection search in $\rho$ (at any given fixed $T$),
repeatedly solving Eq.~\eqref{mctf} numerically until the
transition point $\rho_c(T)$ is located with the desired accuracy. To this
end, wave
vectors are discretized to a grid of $M$ points with spacing $\Delta q$.
The results presented below are typically obtained for $M=315$ and
$\Delta q=0.4$, with values as large as $M=600$ ($\Delta q=0.04$) for the
SWS. $S(q)$ was obtained as explained above, on a grid with spacing
$\Delta\rho=0.01\,\text{nm}^{-3}$ for each temperature and interpolated linearly between
those points; since $S(q)$ varies smoothly, this interpolation leads to
negligible errors. The results were checked for representative points with finer $q$ ($\Delta q = 0.2$) and $\rho$ ($\Delta\rho=0.001\,\text{nm}^{-3}$) grids, and with larger ranges ($M=420$).

Each point on the MCT transition line is characterized by a real number
$1/2\le\lambda\le1$, called the exponent parameter, that can be calculated
knowing the coupling constants and $f^c(q)$. For the standard glass
transition there holds $\lambda<1$, and the decay of the density correlation
function in the liquid is, asymptotically close to the transition, governed
by two power laws: $\phi(q,t)-f^c(q)\sim t^{-a}$ for the relaxation towards
the plateau, and $\phi(q,t)-f^c(q)\sim-t^b$ for the initial decay from
this plateau. The exponents $a$ and $b$ are determined by $\lambda$,
\begin{equation}\label{exponent}
  \frac{\Gamma(1-a)^2}{\Gamma(1-2a)}=\lambda=
  \frac{\Gamma(1+b)^2}{\Gamma(1+2b)}\,.
\end{equation}

The case $\lambda\to1$ (corresponding to $a\to0$ and $b\to0$) signals the
breakdown of this power-law asymptotic expansion for $\phi(q,t)$, and points
with $\lambda=1$ are termed higher-order glass transitions. In their
vicinity, the asymptotic shape of correlation functions is better described
by powers of $\log t$ \cite{Goetze2002}. Such higher-order singularities have
been found in the short-range attractive colloidal systems discussed above.
There, they indicate a discontinuous cross-over between two qualitatively
distinct types of glass -- the one driven by repulsive caging, and the one
driven by attractive bonding. The higher-order transition point vanishes
for larger attraction ranges, $\delta>\delta_c$, with $\delta_c\approx0.045$
in the SWS. But even for $\delta$ larger than but close to $\delta_c$,
there remains a region along the glass-transition line $T_c(\rho)$ where
$\lambda$ attains a maximum close to unity and where precursors of the
higher-order transition have been found experimentally or in simulations
\cite{eckert,pham,puertas_a4,zaccarelli_a4}. Hence the approach of $\lambda$ to unity
upon lowering the temperature in a system with attractive interactions can be
taken as indicative for the crossover to an attraction-driven glass.

\section{Results}\label{results}

\begin{figure}
\includegraphics[width=.9\linewidth]{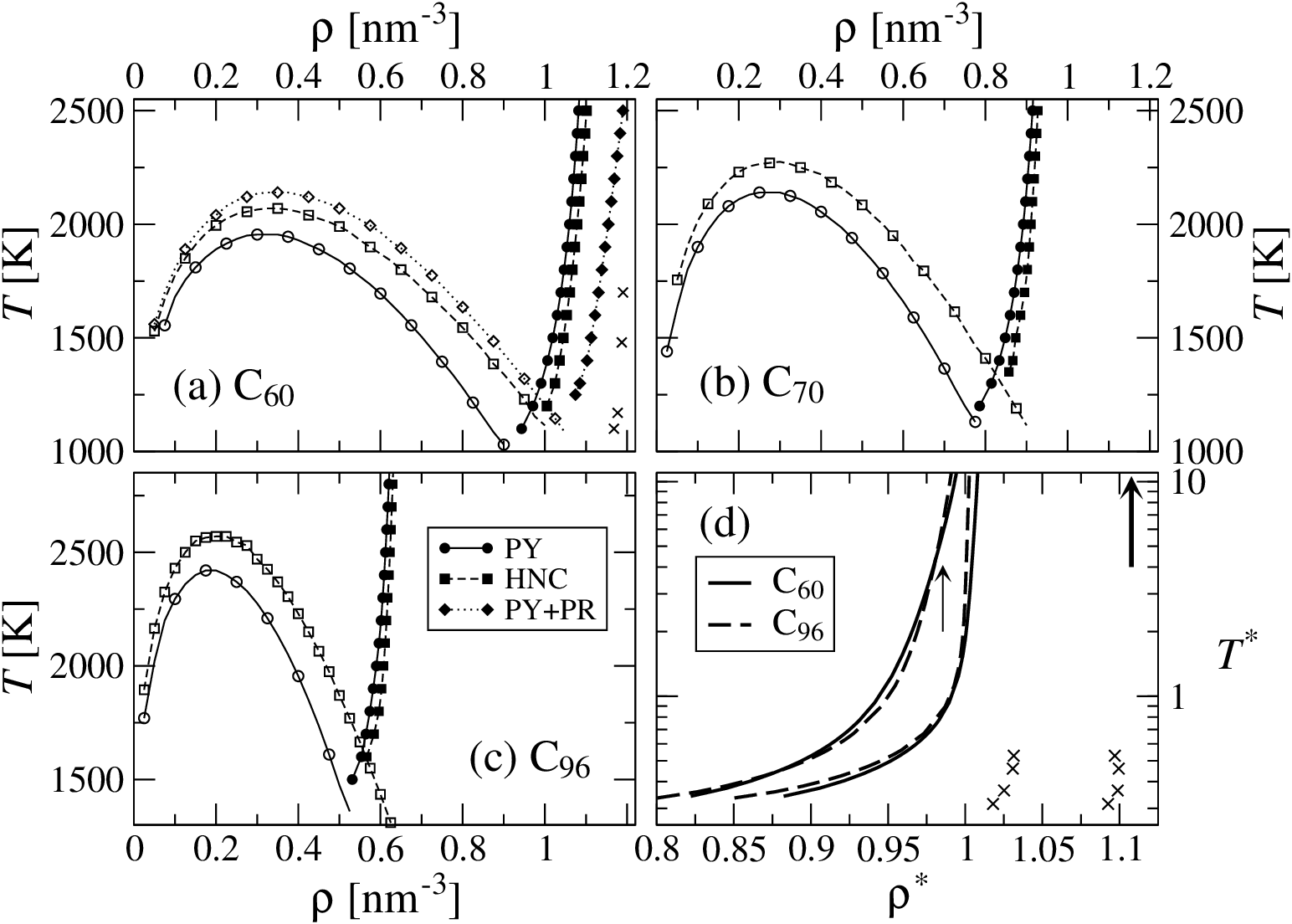}
\caption{\label{cn_pyfig}
  MCT-glass transition lines (filled symbols and lines in panel d)
  for $\text{C}_{60}$, $\text{C}_{70}$, and $\text{C}_{96}$ models:
  in panels (a) to (c), solid (dashed) lines with circle
  (square) symbols are the results for the Girifalco potential using the
  PY (HNC) closure for $S(q)$. Diamond symbols with dotted lines correspond
  to the PR potential for $\text{C}_{60}$. Lines with open symbols indicate
  the loci of points below which no solution for $S(q)$ exists within the
  OZ closure (indication of the gas--liquid spinodal).
  Crosses are MD simulation results from Ref.~\cite{abramo_glass}.
  Panel (d) shows the PY and MD (Ref.~\cite{abramo_glass}) results in terms of dimensionless
  quantities $\rho^*$ and $T^*$ defined according to the BH (left) and WCA (right) prescriptions (see text for details).
  The thin (thick) arrows in panel (d) indicate the $T\to\infty$ transition
  point for hard spheres as calculated from MCT (determined from experiment),
  $(6/\pi)\rho*=0.516$ ($0.58$)
}
\end{figure}

Let us start by discussing the MCT results for the (idealized)
glass-transition lines in the temperature--density plane, $T_c(\rho)$.
Fig.~\ref{cn_pyfig} shows the transition lines for $\text{C}_{60}$,
$\text{C}_{70}$, and $\text{C}_{96}$. Results are shown using both the
HNC (filled squares) and the PY (filled circles) closures to the OZ equation, in order to indicate
the degree of uncertainty imposed on the MCT results by different
approximations to $S(q)$. The respective no-solution boundaries for the
two closures are also shown (HNC: open squares; PY: open circles); they can be taken as a rough
approximation to the gas--liquid spinodal. Note that the binodal, as well as
the freezing and melting lines, being thermodynamic rather than kinetic in
origin, are not shown.

There is a small but noticeable difference between the glass transition curves calculated
within the two different closure approximations. Since HNC and PY can be
seen as limiting cases for a number of more refined closure schemes
\cite{caccamo_rev}, we take the difference between the two transition lines
to be an indication of that error contribution to $T_c(\rho)$ that is not
inherent to MCT. Note that even if the difference is small in terms of
the critical density $\rho_c$, it can still lead to a significant uncertainty
in determining $T_c$, as the lines run nearly vertically in the
$T$-versus-$\rho$ plots shown. The PY-MCT transition is consistently shifted
towards lower coupling strength (smaller $\rho$ and higher $T$) compared to
the HNC-MCT line. This is different from the estimated spinodal regions,
where PY generally estimates lower $T$ than HNC.
This underlines that different aspects of the equilibrium structure are
responsible for the two phenomena.

Fig.~\ref{cn_pyfig} also exhibits results for $\text{C}_{60}$ based on the PR
potential together with the HNC closure approximation [diamonds and dotted
lines in panel (a)]. This considerably softer potential displaces the MCT
glass-transition line to even higher densities. The uncertainty coming from
the potential modeling is much greater than the one due to different OZ
closures. Also included in the figure as crosses are some MD simulation results
\cite{abramo_glass} for $\rho_g(T)$, determined from the nonequilibrium system
following a temperature-quench. As expected (due to the cutting off of the MCT glass transition by hopping processes), $\rho_g>\rho_c$. But the shape of
the two lines is roughly similar.

The fullerene diameter and energy scale both change upon varying $N$, leading
to a known shift of the phase diagram towards lower $\rho$ and higher $T$
\cite{fernandes03,chen03}; the MCT transition line follows this
trend as expected. Panel (d) of Fig.~\ref{cn_pyfig} is an attempt to scale out these broad effects: it shows the transition
lines in reduced units as discussed above, using both the BH and WCA definitions of the effective diameter $\sigma_\text{eff}$. Results for $\text{C}_{70}$
are omitted to avoid overcrowding. The difference between the two rescalings is marked -- use of the WCA-$\sigma_\text{eff}$ shifts the glass transition line to higher $\rho^*$ for all $T^*$. This can be understood by recalling that the WCA definition includes a section of repulsive potential ($V(r)<0$, $\partial_r V(r)>0$) omitted in the BH approach, leading to a higher $\sigma_\text{eff}$. However, the two representations of our results have two major features in common. Firstly, both show a clear bending across of the glass transition lines to lower $\rho^*$ at lower $T^*$, even after scaling out (through $\sigma_\text{eff}$) of the growth in the effective size of the repulsive core as the temperature is decreased. Secondly, the transition lines for $\text{C}_{96}$ are at markedly lower $\rho^*$ than those for $\text{C}_{60}$ at lower $T^*$, even though the different fullerene lines are close (and cross if BH-$\sigma_\text{eff}$ is used) at high $T^*$. Both these points are suggestive of the enhancement of arrest through (relatively) short-ranged attraction in these systems.

However, the form of the glass transition lines at high $T^*$ suggests that we might encounter problems in using these rescaled plots to make firm conclusions about the nature of the glassy behavior. As $T^*\to\infty$, one expects the potential to become more and more
HS-like (since the Girifalco potential has a hard-core excluded
volume contribution), and hence the transition line to approach the
HS value, $\rho_c^*=0.516(6/\pi)$ \cite{mayr_hs} for the MCT-PY calculation.  Similarly, the simulation values for $\rho^*_g$ are expected to approach the
experimentally-determined value for the HS glass transition,
$\rho_g^*\approx0.58(6/\pi)$. Both asymptotic values are indicated in panel
(d) of Fig.~\ref{cn_pyfig} as vertical arrows. 
An ideal choice of effective diameter would collapse the glass transition lines to the HS $\rho_c^*$ (making them vertical) as soon as the temperature became
large compares to the potential depth.
However, both the definitions used here fail to produce any collapse of our MCT results over the temperature ranges considered here, despite the fact that our highest $T^*$ correspond to $T\approx50000\,\text{K}$. In addition, both rescalings move the glass transition lines at high $T^*$ to higher $\rho^*$ than the HS value -- a result of the slow approach of these $\sigma_\text{eff}$ to the HS diameter $\sigma$ referred to in Section \ref{fullerenes}.

Note that in Refs.~\cite{abramo_glass,abramo_glass2}, using the
WCA-$\sigma_{\text{eff}}$, it was concluded that $\rho_g^*\approx0.574(6/\pi)$ is
essentially equal to the hard-sphere transition value and independent of
$T^*$, and hence that the glass formation was entirely repulsion-driven. However, given the lack of a clear-cut physical reason for choosing a particular $\sigma_\text{eff}$, one needs to be careful in drawing conclusions solely from the
numerical values of $\rho^*_g$ defined in terms of such a $\sigma_\text{eff}$ over a limited temperature range. In the following, we will try to find find signatures of attraction- or repulsion-driven glassy behavior that are independent of the choice of reference length-scale. Specifically, we will study the difference between the MCT results computed with the full $V(r)$ and those using the purely repulsive $V_\text{cut}$ employed in the definition of the WCA-$\sigma_\text{eff}$. The exponent parameter $\lambda$ (Eqn.\ \ref{exponent}) and the non-ergodicity parameter $f^c(q)=\lim_{t\to\infty}\phi^c(q,t)$ (given by Eqn.\ \ref{mctf}) will also be calculated, predicting, respectively, the form of the relaxation of density fluctuations and the strength of the arrest.

\begin{figure}
\includegraphics[width=0.8\linewidth]{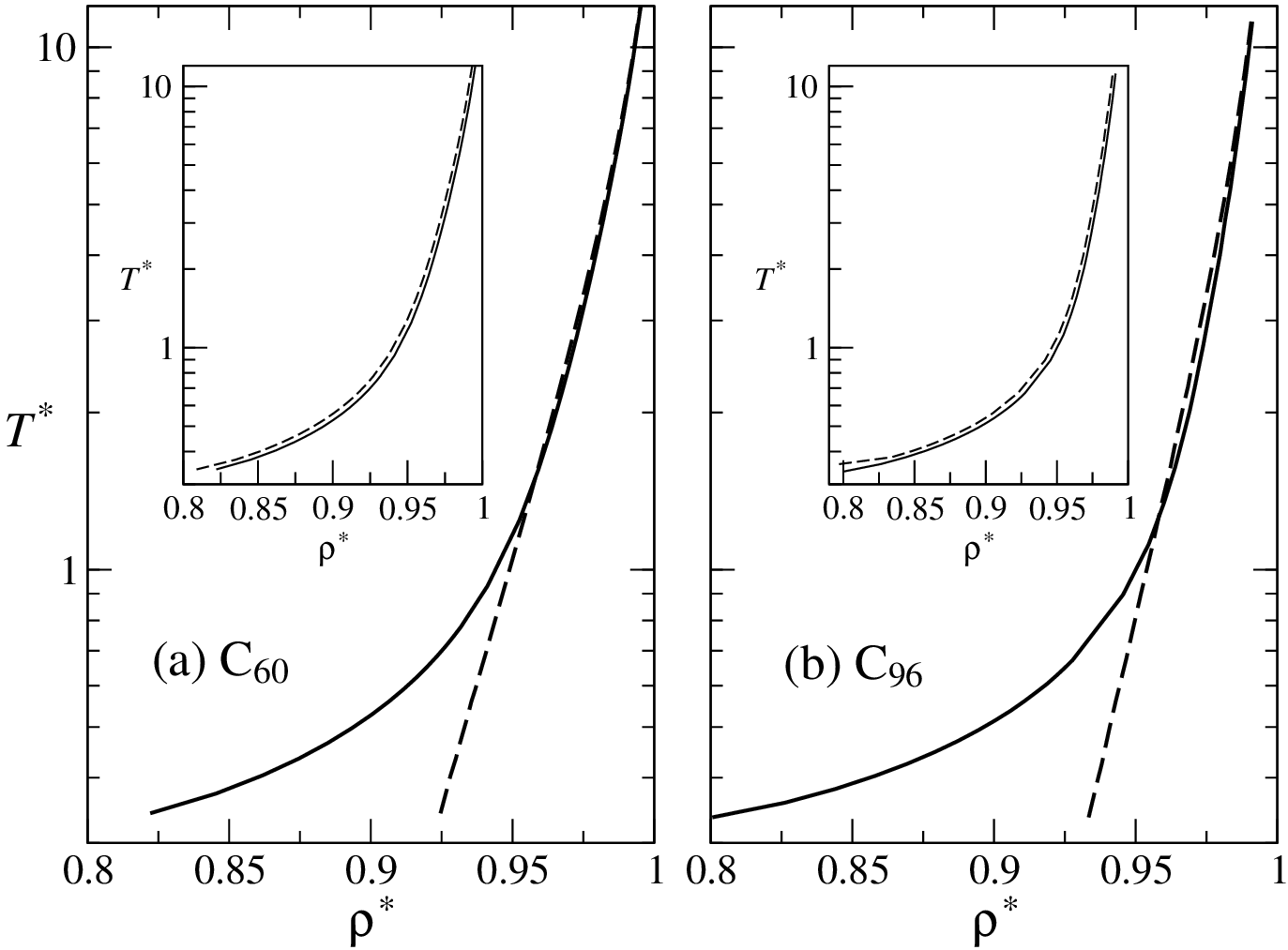}
\caption{\label{attractfig}
  Glass transition lines from MCT with the PY approximation for
  $\text{C}_{60}$ (left panel) and $\text{C}_{96}$ (right panel). The solid
  lines reproduce the transition lines from Fig.~\ref{cn_pyfig}. The dashed
  lines correspond to the analogous results where the Girifalco potential has
  been replaced by its purely repulsive version, $V_{\text{cut}}(r)$,
  Eq.~\eqref{cutoff-potential}. Insets: solid lines are the full-potential
  transition lines, and dashed lines are those obtained from cutting of
  $S(q)$ at low $q$ in the MCT integral, see text for details.
}
\end{figure}

The first of these calculations is aimed at distinguishing the influence of the attractive
part of the Girifalco potential on the glass transition from that of the repulsion. The MCT transition lines are computed for the cut-off
Girifalco potential, Eq.~\eqref{cutoff-potential}, which excludes any
attraction but retains the functional shape of the repulsive core. A similar
comparison of the LJ model with its cut-off variant yields MCT transitions
that are almost indistinguishable from each other \cite{tv}, indicating that
purely attractive effects are negligible for all temperatures.

As shown in Fig.~\ref{attractfig} for $\text{C}_{60}$ and $\text{C}_{96}$, the
situation is different in the Girifalco potential. The glass transition
obtained from the repulsive core of the potential only (dashed lines) only
follows the full-potential result at very high temperatures. It shows a small
deviation in the intermediate-temperature region, and below $T^*\approx1$ it
deviates significantly: it bends over to much lower densities, indicating
that for $T^*\lesssim1$, attraction does matter for the glass transition.

This bend is a genuine kinetic, attraction-driven effect, and in particular
not associated with the proximity of the spinodal. To confirm this, we plot
(insets of Fig.~\ref{attractfig}) a comparison of the full-potential
transition line with one obtained from a model that does have attraction, but
no spinodal effects entering into MCT: the dashed lines in the insets of
Fig.~\ref{attractfig} were obtained by cutting off the $S(q)$ determined from
the full Girifalco potential at $q<q_\text{peak}/3$, thereby eliminating the
sharp increase in $S(q\to0)$ signalling the approach to the spinodal.
The resulting MCT
transition is close to the one including the spinodal, indicating
that the latter plays no important role, and in particular cannot be responsible
for the discussed deviation of the attractive system from the purely repulsive
one. Note that the small shift visible in the insets of the figure is due to
the details of our cutoff procedure which overestimates the MCT coupling
constants.

Based on the above results we can conjecture that the Girifalco-model glass
transition in the experimentally accessible temperature range shows strong
attraction-induced effects. There exists a region at relatively low densities,
$0.8<\rho^*\lesssim0.95$, where the Girifalco model displays a glassy region
that would not be there for a purely repulsive mechanism. We note further that,
also in Fig.~\ref{attractfig}, a ``relative reentry'' can be seen - there is a
region $1<T^*<10$ where the purely repulsive system glassifies as \emph{lower}
densities than the one including attraction. Again, one can allude to the SWS
here, where this phenomenon is observed for $\delta\lesssim0.1\sigma$, albeit
the reentry there is with respect to the hard-sphere transition value. In our
case, the reentry is only relative to the repulsive-transition line; that the
latter is not at constant $\rho^*$ indicates that the Barker-Henderson
effective diameter does not account completely for all soft-core effects.

\begin{figure}
\includegraphics[width=0.8\linewidth]{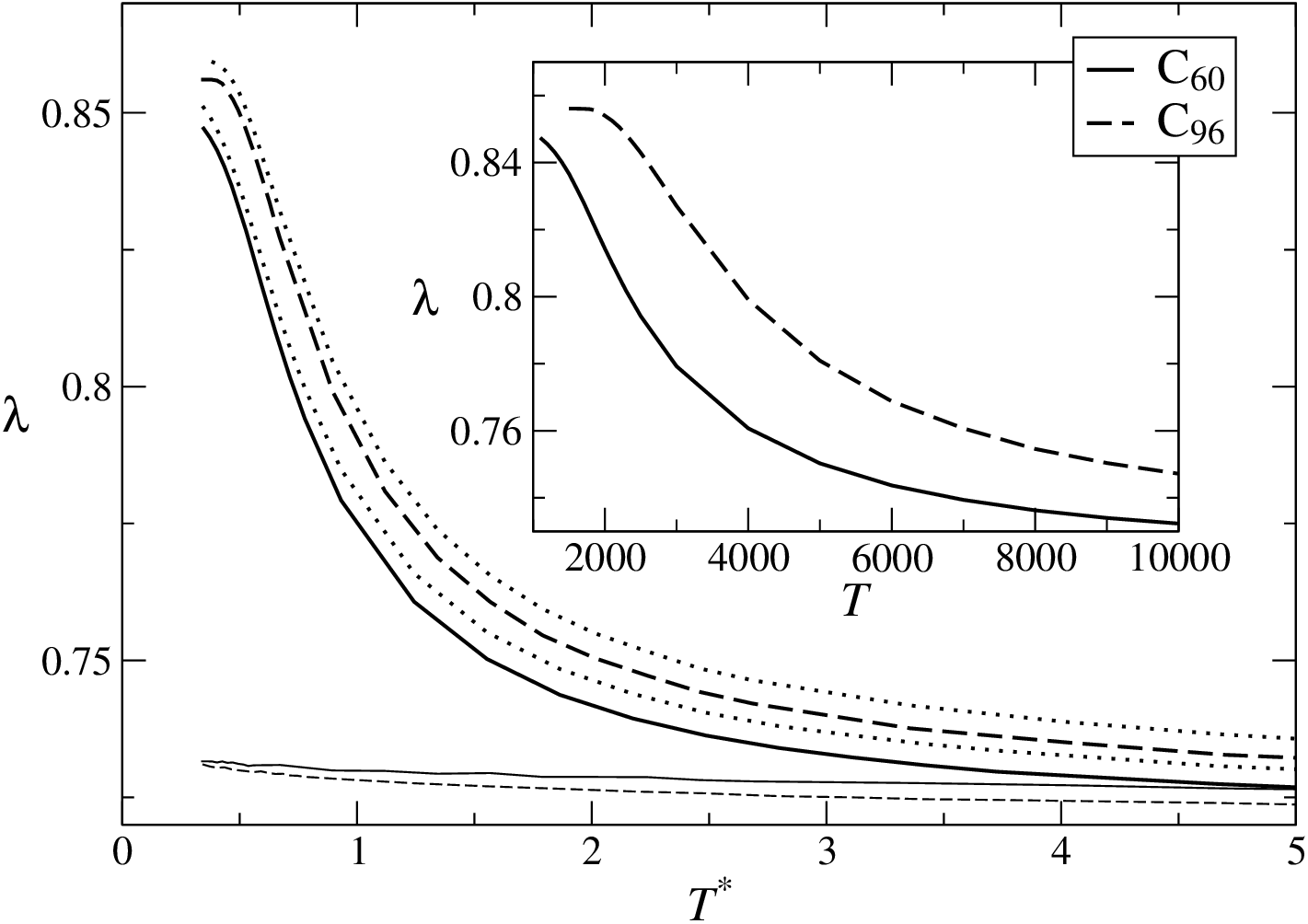}
\caption{\label{lambdafig}
  MCT exponent parameter $\lambda$ evaluated along the glass transition line,
  for $\text{C}_{60}$ (solid lines) and $\text{C}_{96}$ (dashed), as
  function of reduced temperature $T^*$. The upper set of curves corresponds
  to the full potential, while the lower corresponds to the cutoff potential,
  Eq.~\eqref{cutoff-potential}. Dotted lines are results evaluated from the
  full potential without the spinodal low-$q$ region.
  Inset: $\lambda(T)$ plotted as a function of unscaled temperature.
}
\end{figure}

If there is indeed a cross-over from a repulsive glass (at extremely high $T$)
to an attraction-affected one, the MCT exponent parameter $\lambda$ should,
according to the square-well analogy, show a corresponding change as a
function of $T^*$, indicating the cross-over region by a peak.
In Fig.~\ref{lambdafig}, we plot the $T^*$-dependence of $\lambda$ for the
$\text{C}_{60}$ and $\text{C}_{96}$ potentials. Indeed, $\lambda(T^*)$ rises
significantly for $T^*\lesssim1$ and displays a peak at $T^*\approx0.5$ in both
models. This peak is, however, quite close to the region where the glass
transition line terminates at the spinodal, although it is in no way connected
to it. The last statement is again seen from the $\lambda(T^*)$
corresponding to the model with a low-$q$ cutoff in $S(q)$ (see discussion of
Fig.~\ref{attractfig}) being close to the full-$S(q)$ calculated
one (the small rise in $\lambda$ being due to the overestimation of the MCT coupling constants referred to earlier). In contrast, the purely repulsive systems according to
Eq.~\eqref{cutoff-potential} display a value $\lambda\approx0.73$ that is
essentially independent of $T^*$ and consistent with the values one gets for
the Lennard-Jones or hard-sphere systems.
Again, this clearly demonstrates the influence of attraction at sufficiently
low temperatures. $\lambda$ however remains bounded by $0.85$ ($0.86$) for
$\text{C}_{60}$ ($\text{C}_{96}$), hence no higher-order glass-transition
singularity is predicted for these models. The peak in $\lambda(T^*)$
increases somewhat with increasing $N$, i.e., with decreasing attraction
range, so that one can conjecture the existence of a higher-order singularity
in giant fullerene systems. But unrealistically large $N$ would need to be
considered for this.

As explained in Sec.~\ref{mct}, $\lambda$ determines the exponents for the
asymptotic description of the time-dependent relaxation functions. To the
``standard'' value $\lambda=0.73$ correspond exponents $a=0.591$ and
$b=0.314$. In the $\text{C}_{60}$ model, the values drop to $a\approx0.396$
and $b\approx0.250$ at the maximum in $\lambda$; for $\text{C}_{96}$ we
similarly get $a\approx0.381$ and $b\approx0.244$. However, these figures are still
significantly different from zero, so that logarithmic decay is probably not
observable in fullerene glass formers.

\begin{figure}
\includegraphics[width=0.8\linewidth]{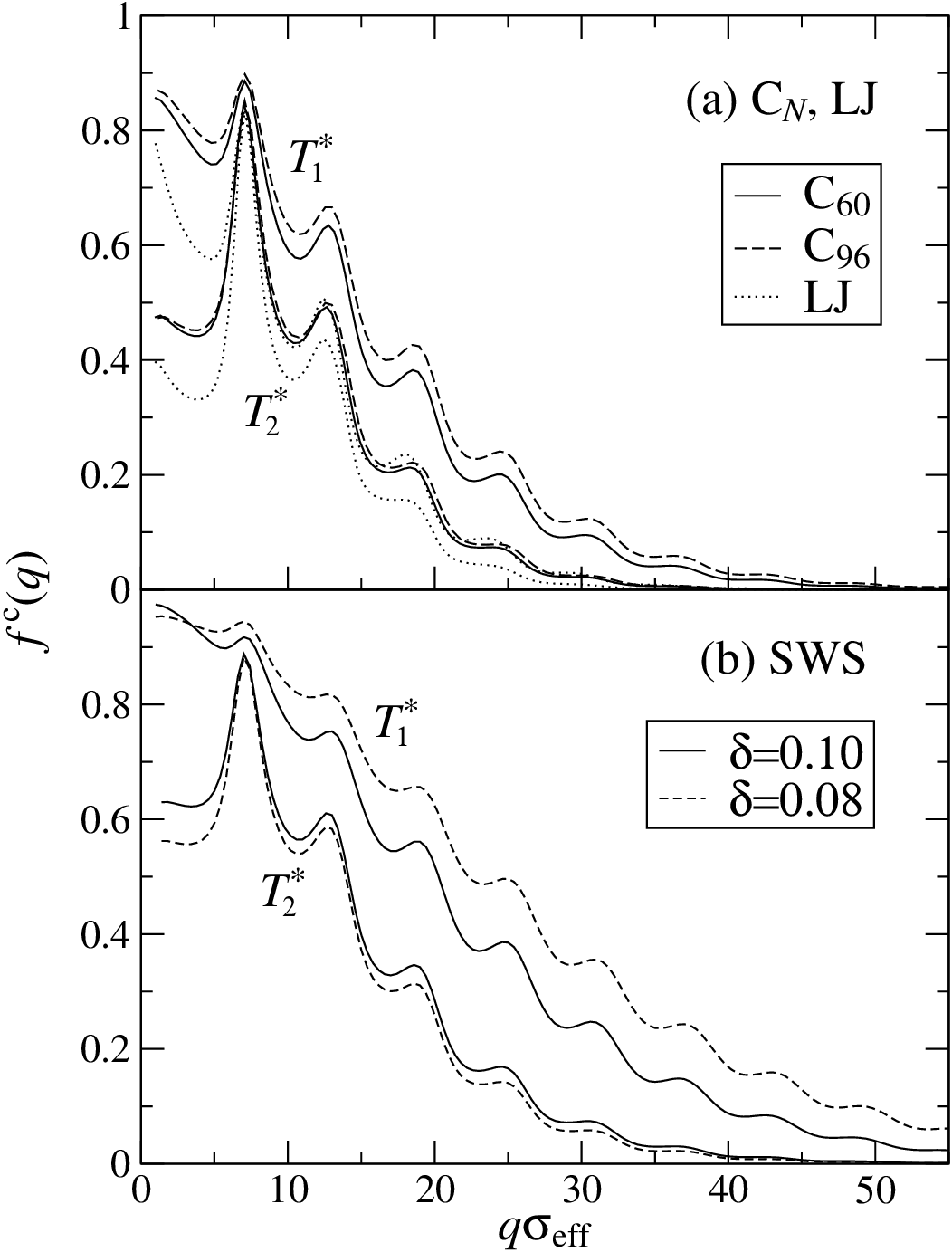}
\caption{\label{fqcombfig}
  Nonergodicity parameters $f^c(q)$ at the MCT transition as functions
  of rescaled wave vector, $q^*=q\sigma_{\text{eff}}^3$.
  Upper panel: $\text{C}_{60}$ (solid lines) and $\text{C}_{96}$ (dashed);
  upper curves correspond to $T^*_1=0.5$, lower curves to $T^*_2=2$
  (PY approximation). The dotted lines show the corresponding results for the
  Lennard-Jones system.
  Lower panel: $f^c(q)$ for the square-well system, with attraction range
  $\delta=0.10$ (solid lines) and $\delta=0.08$ (dashed), at temperatures
  $T^*_1\approx0.22$ and $T^*_2=0.5$ (MSA approximation).
}
\end{figure}

As mentioned above, attractive and repulsive glasses can be distinguished by
their localization length, and similarly by the shape of the $f^c(q)$
quantifying the degree of arrest. Figure~\ref{fqcombfig} shows these plotted
as functions of rescaled wave vector, $q^*=q\sigma_{\text{eff}}^3$, for
$T^*=0.5$ (roughly corresponding to $T\approx1500\,\text{K}$) and $T^*=2$.
A comparison with the corresponding LJ result (shown as dotted lines)
reveals that the Girifalco model predicts a higher degree of arrest than that occurring in
the Lennard-Jones liquid: the $f^c(q^*)$-versus-$q^*$ curves for $\text{C}_N$
are systematically above the LJ ones. In particular, there are two noteworthy
trends exhibited by these curves, one upon changing temperature, the other
upon changing $N$.

For the LJ system, $f^c(q^*)$ changes relatively little with $T^*$, except for the
low-$q$ region as the spinodal is approached. This increase in the $q\to0$
value is connected with the divergence of compressibility there. For the
Girifalco models, $f^c(q^*)$ changes rather more with temperature. Again, the
fact that attraction plays an important role in the formation of fullerene
glasses whereas it does not in LJ glasses, manifests itself here.

Shortening the attraction range with increasing $N$, we find the $f^c(q^*)$ to
increase at fixed $T^*$, indicating that particle caging is enhanced.
Extracting a measure of the localization length from these curves,
one concludes that this length shrinks with shrinking width of attraction.
However, from the half-width of the $f^c(q^*)$-versus-$q^*$ curve
one still estimates localization lengths of the order of $5\%$ for
the systems studied here, rendering them intermediate between the LJ reference
system and the truly short-ranged SWS. To highlight the connection to this
SWS, we show in the lower panel of Fig.~\ref{fqcombfig} similar $f^c(q^*)$ for
two such systems, $\delta=0.1$ and $\delta=0.08$. This leads to localization lengths that are approximately those observed in the Girifalco system. Also, the values of $\lambda$ are similar: $\lambda\approx 0.81$ for $\delta =0.08$ and $\lambda\approx 0.79$ for $\delta=0.10$ at $T_1^*$, to be compared with $\lambda\approx 0.83$ for $\text{C}_{60}$ (see Fig.\ \ref{lambdafig}).

Again, as $T^*$ is lowered, a cross-over occurs from relatively narrow
$f^c(q^*)$-versus-$q^*$ curves to significantly
wider ones, signalling the crossover from a repulsive to an
attraction-driven glass. Note that the numerical values for $T_1^*<T_2^*$ cannot be
compared immediately, since the SWS data has been obtained in the
mean-spherical approximation (MSA) to $S(q)$, which results in an intrinsically
different $T^*$-scale \cite{dawson}.
In the SWS data shown, the crossover is more evident than it is in the
Girifalco model. Note that neither of the two SWS are short-ranged enough to
exhibit a higher-order glass transition.

The comparison with the SWS leads quite naturally to the question how to
define an effective attraction range $\delta_{\text{eff}}$ for smooth
potentials such as the Girifalco one.
One such definition has been proposed by Noro and Frenkel \cite{NoroFrenkel}
and is based on an extended law of corresponding states: one compares
systems with smooth interactions to a SWS at a corresponding state (in terms
of rescaled density and temperature), for which also the second virial
coefficient relative to that of a HS system,
$B_2^*$, is matched. Since $B_2^*$ depends on the attraction range,
this gives a prescription to determine an $\delta_{\text{eff}}$.
From the Noro/Frenkel mapping we read off $\delta_{\text{eff}}\approx0.14$
for the $T^*\approx0.5$ discussed in Fig.~\ref{fqcombfig}. However, it appears that the values of $\lambda$ correspond more closely to a SWS with a
narrower range, $\delta\lesssim0.10$. This can be rationalized easily: the
mapping by Noro and Frenkel emphasizes a correspondence of $B_2^*$, which
is argued to take on roughly constant values close to the liquid--gas critical
point. The physical mechanisms responsible for the change in $f^c(q)$, on the
other hand, are quite different from the spinodal effects, as pointed out in
connection with Figs.~\ref{cn_pyfig} and \ref{attractfig}. There is no reason
to expect that a $\delta_{\text{eff}}$ suitable for mapping the spinodal region
will be the most useful one to map the glass-transition region to that of a
SWS.

\section{Conclusions}\label{conclusion}

We have calculated idealized glass transition lines for a number of fullerene
systems with various sizes, using the mode-coupling theory of the glass
transition and an effective pair-potential description. Our results are
readily testable using molecular-dynamics computer simulation, such as
performed recently \cite{abramo_glass,abramo_glass2,ruberto}.

Glass transition lines are found for densities of the order of
$1\,\text{nm}^{-3}$ for $\text{C}_{60}$, bending over to slightly lower
densities at temperatures below the gas--liquid critical point. The results
for $\text{C}_{70}$ and $\text{C}_{96}$ are qualitatively similar, but shifted
in the temperature--density plane according to the change in natural energy
and length scales of the different systems. At least within MCT, these glass
transitions, being purely kinetic in origin,
appear completely unrelated to the gas--liquid spinodal or similar
thermodynamic questions, and hence
to the longstanding question whether fullerene liquids are stable or only
metastable with respect to sublimation.
Note also that a
common way to suppress crystallization and hence study the metastable
liquid in colloidal suspensions is to make use of their polydispersity.
It might be interesting whether
a similar polydispersity arising in the production of fullerenes \cite{sloan}
could play the equivalent role, something that has already been
indicated in simulation studies of
binary fullerene mixtures \cite{ruberto}.
Finally, in the observation of our predicted MCT transition line, one will
need to consider (in analogy to standard molecular liquids) the problem of
additional (hopping) relaxation processes, which may cutoff the MCT transition
as such, to different extent at different temperatures. However, the
observability of our predictions should be not worse than in other molecular
liquids, where MCT has been applied with great success at least in a limited
region on the liquid-side of the transition \cite{goetze_exp}.

The glass transitions we have discussed are strongly influenced by
inter-particle attraction, which is, at least within MCT, a clear contrast to
standard molecular
glasses where interactions are well described by Lennard-Jones type potentials.
This attraction-domination leads to an occurrence of the glass transition at
lower densities than in the purely repulsive system. It manifests
itself in several ways, in particular through MCT's exponent parameter
$\lambda$ and the plateau values of the time-dependent correlation functions
(or the glass form factors). The exponent parameter, $\lambda\gtrsim0.85$ at
typical temperatures, is found to be significantly higher than in
LJ glasses ($\lambda\approx0.73$), leading to lower exponents $a$ and $b$
for the asymptotic description of the time-dependent relaxation in terms of
power laws.
It should be possible to observe this difference, for example by an
asymptotic analysis of correlation functions measured in MD simulations
\cite{goetze_exp}. Also, the asymptotic form of the divergence of
relaxation times or the diffusivity close to $T_c$ is governed by these
exponents, $\tau\sim|T-T_c|^{-\gamma}$ with $\gamma=1/(2a)+1/(2b)$.
Note that $\gamma\approx 2.44$ in the hard-sphere system, while our
calculations predict $\gamma\approx 3.46$ for the attraction-affected part of
the fullerene glass transition.
The final decay of the correlation function is also often fitted with a
stretched exponential law, $\phi(q,t)\approx A(q)\exp[-(t/\tau(q))^\beta(q)]$,
with some $A(q)\le f^c(q)$ and a stretching exponent $\beta(q)<1$. For
$q\to\infty$, $\beta(q)\to b$ \cite{fuchs_kww}; thus $b$ can be taken as
a measure of the `stretching' exhibited by the correlation functions: low
values of $b$ will correspond to a more stretched decay. It is conceivable
that such an analysis of dynamical correlation functions will reveal a much
broader relaxation spectrum than in usual glasses.
Such tests have, to our knowledge, not been performed yet.

As a second signature,
the glass form factors $f^c(q)$ within the Girifalco model are noticeably
higher than for corresponding LJ states: the fullerene glass is predicted to
be relatively stiff, featuring relatively high plateaus in the dynamical
two-step relaxation process over a wider wave-vector range than usually
observed in molecular glasses: in the LJ system, this plateau has basically
dropped to zero for $q\sigma_{\text{eff}}\approx 30$, while we calculate
$f^c>0.1$ still at this wave vector for the Girifalco system. There, one
needs to reach wave vectors $q\sigma_{\text{eff}}\approx 50$ before the
amplitude of the final relaxation process vanishes.
From the trend observed in $f^c(q)$, we argue that for
$\text{C}_N$ glasses, the localization length (a measure of the average cage
size) should scale with $1/N$, i.e., with the inverse size of the fullerene
molecule and therefore its relative attraction range. This scaling is similar
to that observed in attraction-driven square-well systems, and opposite to
that in the standard class of repulsive systems. Comparison of our predictions
for $f^c(q)$ with simulations or experiment would require a rescaling of
density \cite{vanmegen} and probably also temperature \cite{sperl_isodiff} as
a result of the tendency of MCT to underestimate the coupling required to
produce arrest \cite{sperl_isodiff,bergenholtz_langmuir}. It is possible that the conclusion
of Abramo {\em et al} \cite{abramo_glass, abramo_glass2} that the glass transition in $\text{C}_{60}$ is
hard-sphere-like (from the collapse of their data to a vertical line using the
WCA-$\sigma_{\text{eff}}$) is a result of the fact that the molecular dynamics
transition is shifted to lower temperatures (higher couplings) than that in MCT. Lower temperatures and densities (if accessible) might then be required to see the effects of attraction.

Note that $f^c(q)$ is usually predicted from MCT with much better
accuracy than e.g.\ $T_c$, leading to often quantitatively correct results.
Measurements of $f^c(q)$ would therefore be highly desirable, allowing both a
test of MCT and an estimate of its error in terms of density and temperature.

Attraction-dominated in the sense discussed above, however, does not imply the
phenomena of \emph{very short-ranged} strong attraction, as they have been
discussed in the field of colloid-polymer mixtures. Their hallmarks are
logarithmic decays in the time-dependent relaxation; in this respect, the
fullerene systems we have studied are only intermediate in terms of their
attraction range. However, attempting to map the results to those obtained for
square-well systems can still prove useful, as our comparison of $f^c(q)$
results indicates.

In particular, they can be used to define an effective attraction range of the
fullerene (or similar) systems. We want to emphasize that such a definition is
not unique, and might well depend on the physical problem one is interested
in. Previous discussions have argued mainly in terms of a nearly vanishing
stable-liquid pocket, emphasizing the metastability of the liquid--gas
spinodal as a signature of short-ranged attractions. If one, however, is
interested in high-density kinetic phenomena, the underlying physical
mechanisms mediated by the attraction have little in common with those in the
vicinity of the spinodal, and hence a different measure of
effective attraction range needs to be used.
Such a measure could be based on demanding equality between the $f^c(q)$,
or, less stringently, the localization length at the glass transition.
If one is concerned with glassy dynamics, the best measure might be to
introduce a ``law of corresponding glasses'' based on the MCT exponent
parameter, demanding $\lambda(T*)=\lambda_{\text{SWS}}(T^*,\delta_{\text{eff}}
(T^*))$ to determine $\delta_{\text{eff}}$.

\begin{acknowledgments}
We thank M.E.~Cates for valuable comments. This research was funded by
EPSRC grant GR/S10377. Th.V. acknowledges financial support through DFG
grant Vo~1270/1-1.
\end{acknowledgments}


\end{document}